\documentclass[12pt,preprint]{aastex}
\usepackage{flushrt,float}

\slugcomment{Revised version submitted to the Astrophysical Journal Letters}

\shortauthors{Limongi, Chieffi, Bonifacio} \shorttitle{HE0107-5240}

\begin{document}
 
\title{On the origin of HE0107-5240, the most iron deficient star presently known}

\author{Marco Limongi\altaffilmark{1,2,3}, Alessandro Chieffi\altaffilmark{4,2,3} and Piercarlo Bonifacio\altaffilmark{5}}

\altaffiltext{1}{INAF - Osservatorio Astronomico di Roma, Via Frascati 33, I-00040, Monteporzio Catone, Italy;
marco@mporzio.astro.it}

\altaffiltext{2}{School of Mathematical Sciences, P.O. Box, 28M, Monash University, Victoria 3800, Australia}

\altaffiltext{3}{Centre for Astrophysics and Supercomputing, Swinburne University of Technology, Mail Number 31,
P.O. Box 218, Hawthorn, Victoria 3122, Australia}

\altaffiltext{4}{Istituto di Astrofisica Spaziale e Fisica Cosmica (CNR), Via Fosso del Cavaliere, I-00133, Roma, Italy;
achieffi@rm.iasf.cnr.it}

\altaffiltext{5}{INAF - Osservatorio Astronomico di Trieste, Via Tiepolo 11, I-34131, Trieste, Italy;
bonifaci@ts.astro.it}

\begin{abstract} 

We show that the "puzzling" chemical composition observed in the extremely metal 
poor star HE0107-5240 may be naturally explained by the concurrent pollution of at 
least two supernovae. In the simplest possible model a supernova of quite low mass 
($\rm \sim 15~M_\odot$), underwent a "normal" explosion and ejected $\rm \sim 
0.06~M_\odot$ of $\rm ^{56}Ni$ while a second one was massive enough ($\rm \sim 
35~M_\odot$) to experience a strong fall back that locked in a compact remnant all 
the carbon-oxygen core. In a more general scenario, the pristine gas clouds were 
polluted by one or more supernovae of relatively low mass (less than $\rm \sim 
25~M_\odot$). The successive explosion of a quite massive star experiencing an 
extended fall back would have largely raised the abundances of the light elements 
in its close neighborhood.

\end{abstract}

\keywords{nuclear reactions, nucleosynthesis, abundances -- stars: evolution -- stars: interiors -- stars: supernovae }


\section{Introduction}

Extremely metal poor stars (EMPS) are formed in the very early epochs of Galaxy 
formation by gas clouds chemically enriched only by the very first stellar 
generation. Hence they preserved up to the present time the fingerprints of this 
primordial population. Up to now, more than a dozen of stars having $\rm [Fe/H]\leq 
-3$ (here $\rm [A/B] = Log_{10} (X_A/X_B) - Log_{10} (X_A/X_B)_\odot$, where $\rm 
X_A$ and $\rm X_B$ are the abundances of elements A and B respectively) have been 
discovered, for which a quite large number of element abundance ratios have been 
reliably determined (Bonifacio et al. 1998, Norris, Ryan \& Beers 2001, Depagne et 
al. 2002, Christlieb et al. 2002, Aoki et al. 2002, Carretta et al. 2002, Fran{\c 
c}ois et al. 2003). The picture that emerged from the available data prior to the 
work of Fran{\c c}ois et al. (2003) was the existence of a quite large star to star 
[el/Fe] scatter among the most metal poor stars (McWilliam et al., 1995; Ryan, 
Norris and Beers, 1996); this was interpreted as an evidence of large 
inhomogeneities in the first chemical enrichment of the pristine material. On the 
contrary the new high resolution spectra of four EMPS analyzed by Fran{\c c}ois et 
al. (2003) have shown that there is no evidence of a strong star to star scatter 
for the elements Na to Ti in these stars. This would imply a more "homogeneous" 
enrichment of the pristine material. By combining all the available data (and 
including the C rich stars!), it is possible to identify essentially two groups of 
EMPS. A first one includes stars sharing a) a quite similar abundance pattern for 
all the observed elements from C to Ni and b) (el/Fe) ratios not largely deviating 
from the solar ones: these stars (like, e.g., $\rm CD - 38^{o}245$, $\rm CD -
24^{o}17504$) are naturally interpreted in {terms of
the ejecta  of a single supernova} 
(Chieffi \& Limongi 2002). Stars belonging to the second group (like e.g. CS22949-
037, CS22498-043 and CS22957-027) show a) large enhancements of the light elements 
relative to iron, b) significant star to star abundance scatter and c) an abundance 
pattern of the heavier elements (Si and above) very similar to the one shown by the 
stars belonging to the first group. These stars cannot be explained by the ejecta 
of a "standard" single Supernova.

HE0107-5240 is the most iron deficient star presently known (Christlieb et al. 
2002). Beyond its iron deficiency (1/200000 that of the Sun, i.e. [Fe/H]=-5.3) it 
shows enormous C, N and Na enhancements (by a factor of $10^4$ , $10^{2.3}$ and 10, 
respectively, relative to iron) while Mg, usually found enhanced in the EMPS, is 
almost at the level of iron. This star is the most extreme example of stars 
belonging to the second group and its understanding is of vital importance to 
shed light on the nature of this variegate ensemble of "atypical" stars that are 
preserving, up to the present time, clues about some still unknown characteristic, 
dynamical and/or evolutionary, of the very first stellar generation. 

A discussion of all the possible scenarios that can produce the chemical composition 
observed on the surface of this star is beyond the purposes of the present paper. 
However we wish to briefly comment on the self-pollution scenario that has been 
invoked (Christlieb et al. 2002 and references therein) to explain the large 
overabundances of C and N observed in HE0107- 5240. In this scenario, the  observed 
overabundances of C and N are due to the penetration of the He convective shell 
that forms at the He-flash in the H rich envelope. At the beginning the 
protons engulfed in the He convective shell react with the C produced by the 
$3~\alpha$ (producing therefore a large amount of primary N). Then the energy 
released by the burning of these protons induces a large expansion and cooling of 
all the H-rich mantle (down to the ignition point of the He flash{\bf ).} As a 
result of such an expansion, the He convective shell and the convective envelope 
merge so that a strong dredge up of C and N to the surface occurs. A paper fully 
devoted to the analysis of this scenario is in preparation and it is based on an 
extended set of new and detailed computations extending from the pre main sequence 
up to the asymptotic giant branch (Picardi et al. 2003, in preparation). 
Here we can anticipate that this (self-
pollution) scenario is {\it incompatible} with the observed chemical composition 
of HE0107- 5240. There are at least four different and independent reasons for 
this: 1) the models predict (very robustly) a C/N ratio of the order of 1 while the 
observed ratio is of the order of 140, 2) the predicted $\rm ^{12}C/^{13}C$ ratio 
is 4 while the observed lower limit is 30, 3) the time spent by the star on the Red 
Giant Branch (RGB) after the dredge-up (second RGB) is 1/100 of the time spent on 
the RGB before the dredge-up (this means that one expects roughly 100 RGB zero 
metallicity stars not showing these huge enhancements for each star similar to 
HE0107-5240: they are not observed), 4) the surface gravity of this star is 
incompatible (too large) with the range of values it should have if it were on the 
second RGB. 

\begin{figure} 
\plotone{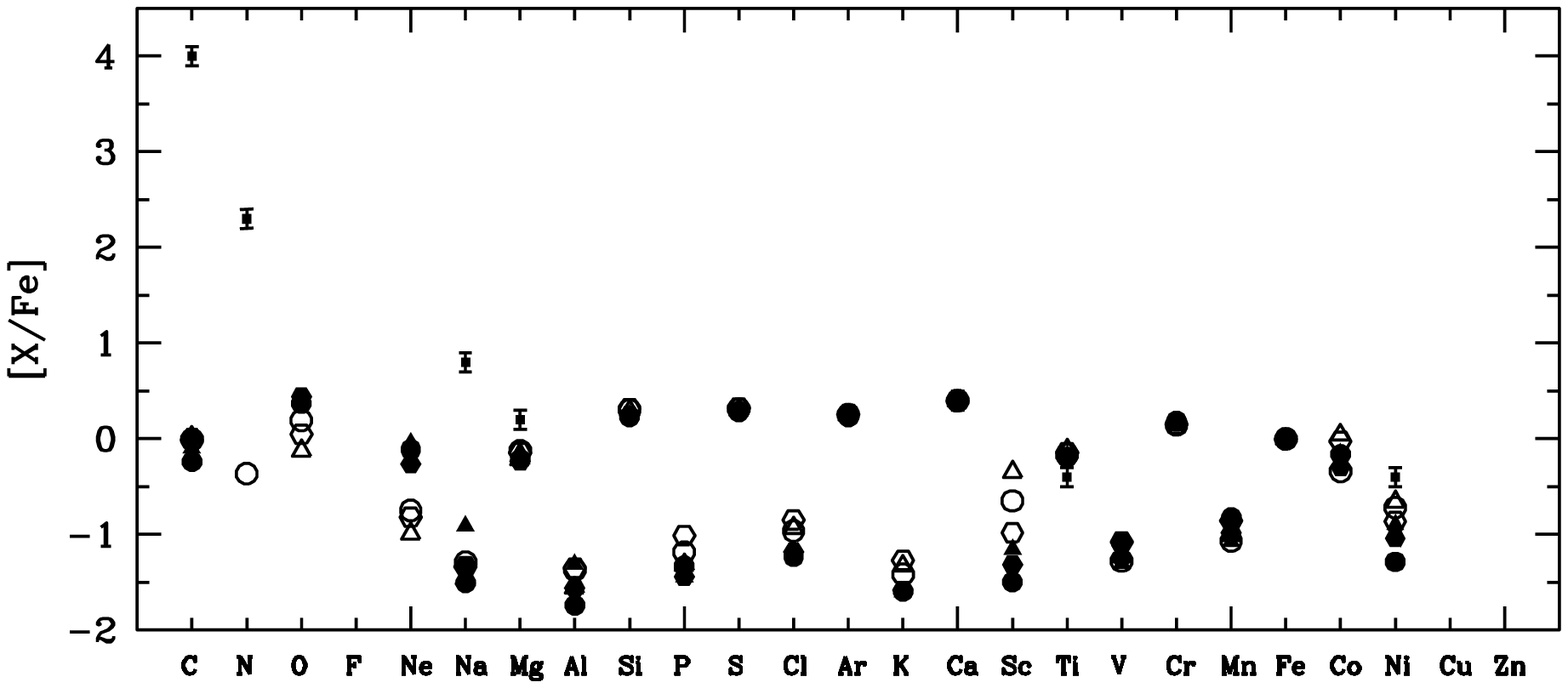}
\caption{
Comparison between the [X/Fe] observed in HE0107-5240 (filled squares) and the ones
predicted by the explosion of stellar models of different masses:
$\rm 15~M_\odot$ ({\em open triangles}),
$\rm 20~M_\odot$ ({\em open hexagons}),
$\rm 25~M_\odot$ ({\em open circles}),
$\rm 35~M_\odot$ ({\em filled triangles}),
$\rm 50~M_\odot$ ({\em filled exagons}) and
$\rm 80~M_\odot$ ({\em filled circles}).
For each model the mass cut has been chosen to fit the observed [Ca/Fe].
\label{inconsistencies}}
\end{figure}

The interpretation of the abundance pattern of the EMPS in terms of ejecta provided 
by a single "standard" primordial core collapse supernova (SNII) fails miserably in 
the case of HE0107-5240 because of two major inconsistencies that show up when 
comparing the models to the observed abundances. The first one can be easily 
understood by reminding that Fe (mainly produced as the unstable $\rm ^{56}Ni$ 
nucleus) is synthesized by both the incomplete explosive Si burning (at 
temperatures in the range $\rm 4<T/10^{9}<5~K$) and by the deeper complete 
explosive Si burning (at temperatures larger than $\rm 5\cdot 10^{9}~K$), while Ti 
and Ni are synthesized only by the complete explosive Si burning. The observed 
abundances of Ti and Ni in HE0107-5240 clearly point towards a mass cut (the mass 
coordinate that separates the ejected matter from the compact remnant) deep enough 
to allow the ejection of at least part of the products of the complete explosive Si 
burning. Conversely, the large overabundance of, e.g., C relative to Fe definitely 
would point towards a mass cut so external that all the layers exposed to the 
complete explosive Si burning were locked in the remnant as well as most of the 
matter exposed to the incomplete explosive Si burning. Figure \ref{inconsistencies} 
clearly shows the inconsistency between the models and the observed abundances of 
the light elements, once the mass cut has been chosen to match the iron peak 
nuclei. 

The second inconsistency concerns the relative abundances among the light elements 
C, Na and Mg. The [C/Mg] and [Na/Mg] ratios observed in HE0107-5240 are +3.8 and 
+0.6 respectively and are incompatible with any mass cut internal to the CO core. 
In other words C, Na and Mg are produced in the CO core in ratios completely 
different from those seen in this star. Figure \ref{trendlight} shows, as an 
example, the [C/Mg] and [Na/Mg] ratios (thick solid and dashed lines) in a $\rm 
35~M_\odot$ progenitor star as a function of the mass of the remnant. The thin 
lines in Figure \ref{trendlight} show as a reference, in the background, the 
profiles of He, C, Na and Mg within the exploding star. It is evident that the 
values observed in HE0107-5240 are never attained within the CO core.

The only way to get rid of the CO core is to let it collapse into the remnant but 
in this way obviously all the elements produced in the CO core would be lost. 
Extended tests done by changing the mass of the precursor and/or some basic 
uncertainties connected to the computation of the evolution of these stars (like 
the $\rm ^{12}C(\alpha,\gamma)^{16}O$ reaction rate) simply confirm these basic 
inconsistencies. 

The observed ratios of [C/Mg], [N/Mg] and [Na/Mg] are compatible with those 
obtained in a He convective shell engulfing some amount of fresh protons. Indeed, 
the protons ingested in the He shell activate a sequence of reactions that lead to 
the production of N, Na and Mg in the right proportions relative to the C produced 
by the $\rm 3\alpha$ reactions. Such a partial mixing between the He convective 
shell and the overlying H rich mantle is a quite common occurrence in stellar 
models of initial zero metallicity (Woosley \& Weaver 1982, Chieffi et al. 2001, 
Fujimoto et al. 1990) because of the low entropy barrier that develops at the H-He 
interface in these stars. 

Recently Bonifacio, Limongi \& Chieffi (2003) suggested that the chemical 
composition of HE0107-5240 is the result of the superposition of the ejecta of two 
primordial SNII: a first of quite low mass would have been responsible for the 
observed iron peak nuclei while an "almost failed" explosion of a quite massive 
star would have been responsible for the observed light element abundances. In this 
paper we will discuss this model more quantitatively and we will show how it can be 
considered as a specific example of a much wider scenario.

\section{A possible interpretation of the element abundance pattern of HE0107-5240}

\begin{figure}
\plotone{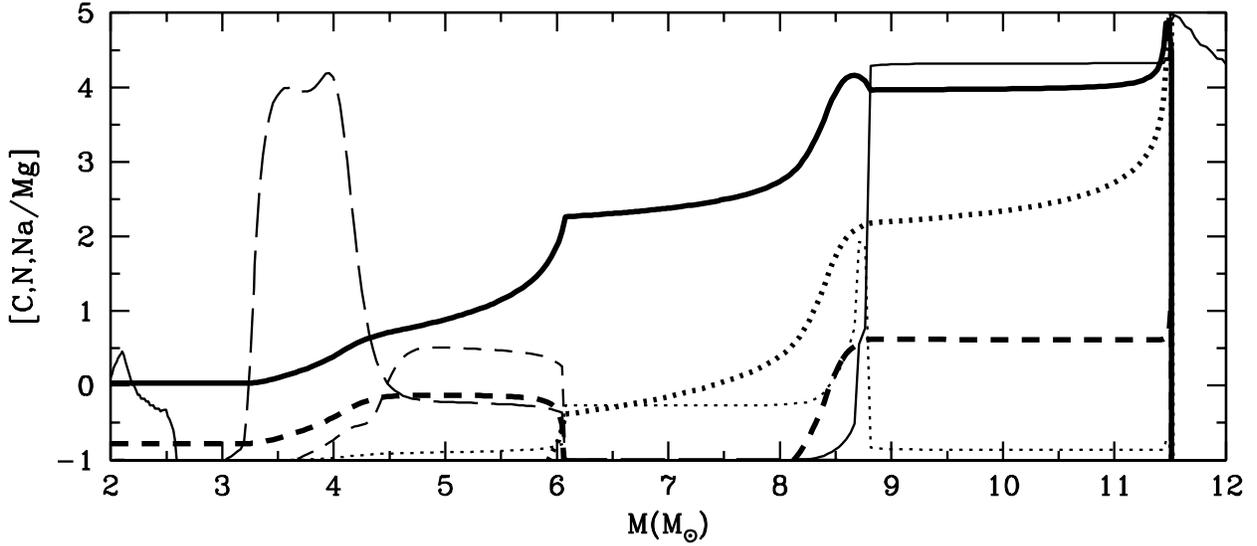}
\caption{ Trends of [C/Mg] ({\em thick solid line}), [N/Mg] ({\em thick dotted 
line}) and [Na/Mg] ({\em thick dashed line}) as a function of the mass of the 
remnant for the $\rm 35~M_\odot$ model. The internal profile of He ({\em solid}), C 
({\em dotted}), Na ({\em short dashed}) and Mg ({\em long dashed}) are shown as a 
reference in background as thin lines. 
\label{trendlight}}
\end{figure}

A thorough analysis of the primordial core collapse supernova models and their associated
explosive yields presented by 
Chieffi \& Limongi (2002) and Limongi \& Chieffi (2002) shows that primordial stars 
of mass of the order of $\rm 35~M_{\odot}$ experience a partial mixing between the 
He convective shell and the H rich envelope and that they produce C, N, Na and Mg 
in this convective shell in relative proportions similar to the ones observed in 
HE0107-5240 (see Figure \ref{trendlight}). A further interesting property of the 
stars in this mass range is that they lie quite close to the mass limit above which 
the explosion almost fails and a very extended fall back onto the remnant occurs 
(Heger \& Woosley, 2002). Hence a primordial core collapse supernova of mass of the 
order of $\rm 35~M_{\odot}$ meets all the requirement necessary to be the star that 
provided the light elements we observe in HE0107-5240.

\begin{figure} 
\plotone{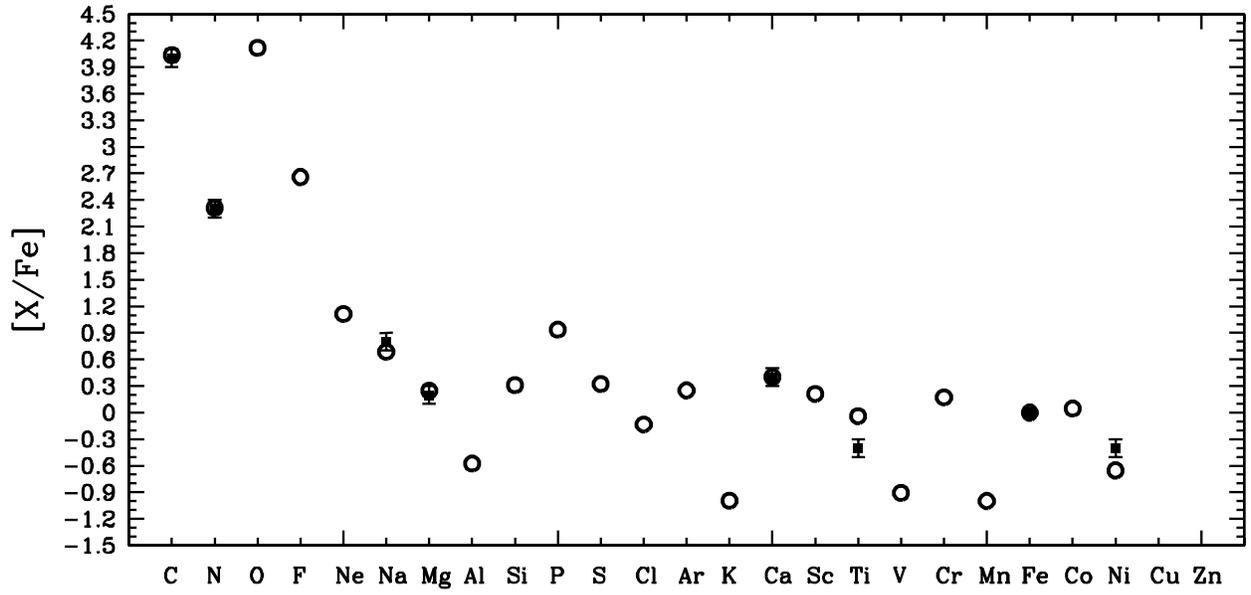}
\caption{
Comparison between the element abundance ratios ({\em open circles}) predicted by the
two supernova model and those observed in HE0107-5240 ({\em filled squares}).
\label{fitfinal}} 
\end{figure}

Since the theoretical [el/Fe] ratios of the elements Si to Ni do not show a clear 
dependence on the initial mass of the precursor, it is not possible to firmly 
identify the star that produced the heavy elements observed in HE0107-5240. This is 
a disadvantage but also an advantage as we will show below. Here, for sake of 
simplicity, let us choose the star that provided the elements produced by the 
explosive burnings. It is quite natural to think that the star responsible for the 
production of the elements Si to Ni had to be significantly less massive than the 
other one. There are two reasons for this: first of all the ejecta of this star 
must not provide enough of light elements (C to Mg) to modify appreciably their 
abundance ratios as provided by the other star. Second, it is reasonable to think 
that the star with the more internal mass cut was also the one with the lowest 
binding energy of the mantle. Both arguments point towards a core collapse 
supernova of quite small mass. The $\rm 15~M_\odot$ star shows up as a good candidate 
for being the star that provided the heavy elements we see in HE0107-5240. In fact, 
by combining the yields provided by these two primordial core collapse supernovae a 
very good fit to HE0107-5240 may be obtained (see Figure \ref{fitfinal}). 

Which was hence the sequence of events that provided the chemical composition we 
observe in HE0107-5240?

First, a star of $\rm 15~M_\odot$ explodes and pollutes the surrounding cloud. The 
fit to the observed [Ca/Fe] requires the ejection of $\rm 5.6\cdot 10^{-2}~M_\odot$ 
of iron, and this fixes the mass of the remnant at $\rm 1.7~M_\odot$. Since the Fe 
mass fraction observed in HE0107-5240 is $\rm 6.7\cdot 10^{-9}$, the ejecta 
provided by the $\rm 15~M_\odot$ star must be diluted in roughly $\rm 7.6\cdot 
10^{6}~M_\odot$ of pristine material in order to obtain the right Fe abundance. 
Within the cloud already polluted by this star, a $\rm 35~M_\odot$ explodes, 
suffers an extended fall back and ejects just the mass above $\rm 9.4~M_\odot$. The 
precise location of the mass cut for this second star is determined by the 
requirement that [C/Mg]=3.8. A last quantity which must be fixed is the mass size 
of the cloud polluted by this second supernova. Since the observed N mass fraction 
in HE0107-5240 is $\rm 1.2\cdot 10^{-6}$ while the supernova ejects $\rm 3.4\cdot 
10^{-4} M_{\odot}$ of N, the ejecta of this star must be diluted with roughly 
$\rm 2.6\cdot 10^{2}~M_\odot$ of gas previously enriched by the smaller supernova. 
Such a "small" mixing of the ejecta produced by this explosion is the natural 
consequence of the extended fall back and of the low velocity of the ejecta. One 
could argue that a $\rm 35~M_\odot$ star evolves faster than a $\rm 15~M_\odot$ star and 
therefore that it should have exploded first. This would be true only in the idealized 
case in which all stars form simultaneously. In practice the lifetime of all these 
stars is short enough (few millions of years) that stochastic inhomogeneities in 
their formation process could have easily led to the explosion of a $\rm 
15~M_\odot$ star before that of a $\rm 35~M_\odot$ star.

It is worth stressing that the final chemical composition predicted by this model 
(shown in Figure \ref{fitfinal}) does not depend critically on the adopted mass cut 
but, instead, it changes slowly and continuously with a change in either of the two 
mass cuts and/or the dilution factors. For example, if the fall back in the $\rm 
35~M_\odot$ star would extend up to $\rm 9~M_\odot$ (instead of 9.4) the predicted 
[C/Fe], [N/Fe] and [Na/Fe] would be 4.0, 2.3 and 0.7 respectively, while a fall 
back extending up to $\rm 11~M_\odot$ (instead of 9.4) would imply a [C/Fe], [N/Fe] 
and [Na/Fe] equal to 3.6, 2.3 and 0.21 respectively. This means that, if this 
scenario is correct, one would expect a wide variety of overabundances of, e.g., 
[C/Fe] and [N/Fe], depending on the amount of fall back experienced by the massive 
star. At variance with respect to  this slow change of the resulting chemical 
composition with varying mass cut in our model, in a scenario like the one 
presented by Umeda and Nomoto (2003), even a tiny change in the mass of the remnant 
would completely change the predicted [el/Fe] ratios by orders of magnitude.

All the discussion presented above naturally points towards a more general scenario. The 
abundance pattern of the elements produced by the explosive burnings does not show 
a strong dependence on the initial mass of the supernova progenitor and hence the 
observed abundances of these elements may well be the result of a single supernova 
explosion but also the combination of two or more supernovae explosions: it is not 
possible to distinguish between these two cases. This also means that the 
similarity in the pattern of the heavy elements observed in the EMPS is neither 
necessarily the consequence of a well mixed environment nor the result of a quite 
constant IMF through the various clouds but it would simply descend from the fact 
that the initial mass does not produce a firm signature on the ejecta of the 
explosive burnings. In these clouds more or less similarly polluted in heavy 
elements, every now and then a quite massive star explodes but, due to the large 
binding energy, suffers an extended fall back and most of the matter (and of the 
energy) remains locked in the remnant. Around these almost failed explosions, low 
mass stars largely enhanced in the light elements may form. Since the ejecta of 
these almost failed supernovae vary slowly with the mass of the remnant, this model 
predicts also a wide spread of possible overabundances with respect to the solar 
values. It is tempting to suggest that at least part of the C-rich
very metal poor stars, which comprise as many as 15\%-25\% of the stars
with [Fe/H]$<-2.5$ (Beers, 1999), are formed in this way.
It goes without saying that this scenario naturally explains the bulk of 
"standard" EMPS as well, because these stars would simply be the ones born far 
from these almost "failed" explosions. They would always have (like they do have) a 
quite similar pattern of the heavy elements but much lower overabundances (and star 
to star scatter) of the light elements relative to Fe.

Figure \ref{fitfinal} shows all the abundances predicted by the specific model 
described above; the measure of any of them in principle will constitute a possible 
test of the present scenario. However, since the predicted element ratios do not 
depend exclusively on the scenario itself but reflect (to some extent) also all the 
assumptions adopted to compute both the pre explosive evolution of the stars (input 
physics and the like) as well as the explosive nucleosynthesis, it would be wise to 
measure as many element abundances as possible to really constrain any scenario.

Let us eventually note that this model also predicts an initial He abundance in 
HE0107-5240 0.015 dex higher than the primordial value and a ratio $\rm 
^{12}C/^{13}C=240$, a value compatible with the lower limit of 30 determined by 
Christlieb et al (2002).

\end{document}